\documentclass[%
 reprint,
 amsmath,amssymb,
 pra,
 superscriptaddress]{revtex4-2}

\usepackage{textcomp}
\usepackage{qcircuit}
\usepackage{braket}
\usepackage{graphicx}
\usepackage{dcolumn}
\usepackage{bm}
\usepackage{bbold}
\begin{document}

\preprint{APS/123-QED}

\title{Computation of molecular excited states on IBM quantum computers using a discriminative variational quantum eigensolver}

\author{Jules Tilly}
\email{jules.tilly@rahko.ai}
\affiliation{Rahko  Limited,  N4 3JP London,  United  Kingdom}
\affiliation{Department of Physics and Astronomy, University College London, WC1E 6BT London, United  Kingdom}

\author{Glenn Jones}
\email{gjones@matthey.com}
\affiliation{Johnson Matthey Technology Centre, Blount's Court, Sonning Common, Reading, RG4 9NH, United Kingdom}

\author{Hongxiang Chen}
\affiliation{Rahko  Limited,  N4 3JP London,  United  Kingdom}
\affiliation{Department of Computer Science, University College London, WC1E 6BT London, United Kingdom}

\author{Leonard Wossnig}
\affiliation{Rahko  Limited,  N4 3JP London,  United  Kingdom}
\affiliation{Department of Computer Science, University College London, WC1E 6BT London, United Kingdom}

\author{Edward Grant}
\affiliation{Rahko  Limited,  N4 3JP London,  United  Kingdom}
\affiliation{Department of Computer Science, University College London, WC1E 6BT London, United Kingdom}

\date{December 24, 2020}

\begin{abstract}
    
	Solving for molecular excited states remains one of the key challenges of modern quantum chemistry. Traditional methods are constrained by existing computational capabilities, limiting the complexity of the molecules that can be studied or the accuracy of the results that can be obtained. Several quantum computing methods have been suggested to address this limitation. However, these typically have hardware requirements which may not be achieved in the near term. We propose a variational quantum machine learning based method to determine molecular excited states aiming at being as resilient as possible to the defects of early Noisy Intermediate Scale Quantum (NISQ) computers and demonstrate an implementation for $\mathrm{H_2}$ on IBMQ. Our method uses a combination of two parametrized quantum circuits, working in tandem, combined with a Variational Quantum Eigensolver (VQE) to iteratively find the eigenstates of a molecular Hamiltonian. 

\end{abstract}

\maketitle

\section{Introduction}

Studies of chemical reactions are currently limited by our knowledge of the total energy and electronic structure of molecules. The electronic state of a molecule can be excited, as in photochemical (catalytic) processes or when characterizing molecules during spectroscopic measurements. The ability to simulate excited electronic structures perfectly would help to identify more efficient reactions and develop new materials \cite{Karim2018, Yazdi2018}.\\
\indent There are inherent limitations to using classical computational techniques to solve for a molecule's excited electronic structure. Molecular system complexity scales factorially in the number of electrons, making even simple molecules difficult to analyse exactly. The advent of quantum computers opens the possibility to address these specific computational challenges with unprecedented accuracy. \\
\indent We propose a hybrid quantum-classical algorithm that uses a quantum machine learning method to find excited states of molecules, building on the ample research already conducted on parametrized quantum circuits~\cite{Mitarai2018,Kandala2017,Mitarai2019,Izmaylov2019,Grant2019, Benedetti2019_2}. From the knowledge of a complex function defining the energy profile of a molecule, we can create an approximate substitute model and then use the Variational Quantum Eigensolver (VQE) to identify the ground state and energy of the molecule \cite{Peruzzo2013}.\\
\indent Typically, resolving energy spectra reduces to the computational task of diagonalizing a matrix representing the Hamiltonian of a molecule. An ideal way to find matrix eigenvectors using a quantum computer is the Quantum Phase Estimation (QPE) algorithm, which could offer an exponential speed-up over the best classical algorithms. It is estimated, however, that proper application of this algorithm would require a fault-tolerant quantum computer. Whitflield et al. \cite{Whitfield2011}, and separately Jones et al.  \cite{Jones2012}, estimate that QPE would require $o{\sim}10^6$ to $o{\sim}10^9$ circuit depth, where depth is defined as the highest number of gate operations that need to be applied to any of the qubits used to implement the algorithm. \\
\indent Variational methods, such as the VQE, are hybrid classical-quantum methods able to operate with lower circuit depth and gates easier to implement. The VQE could prove a valuable tool for approximating the ground state energy of large molecular systems even on Noisy Intermediate-Scale Quantum (NISQ) devices \cite{Preskill2018}. It has already been implemented on many platforms, and displays significantly more resilience to control errors than QPE.\\
\indent The paper is structured as follows. In Section \ref{sec:review} we present an overview of existing methods for computing excited states on quantum computers. We then provide an overview of our method in Section \ref{sec:method}. This is followed by a more detailed derivation of our model in Section \ref{sec:deriv} and a convergence demonstration in Section \ref{sec:demons}. Results for simulations and quantum computer experiments are presented in Section \ref{sec:results} and we provide further details on the quantum computer implementation in Section \ref{sec:Implementation}. Finally we present an analysis of error propagation through several excited states in Section \ref{sec:error_prop}.

\section{Review of existing methods} \label{sec:review}

\indent Several methods have been proposed to calculate excited states using quantum computers. These usually rely on the VQE as a starting point. Most notably, McClean et al. \cite{McClean2017} (see also Colless et al. \cite{Colless2018}) proposed a method based on quantum subspace expansion. Santagi et al. \cite{Santagati2018} proposed a method based on Von-Neumann entropy. Alternative algorithms, tailored to work on NISQ computers, have also been put forward. In particular, Higgott et al. \cite{Higgott2019} developed a variational method in which VQE objectives are minimized concurrent to the overlap between a known ground state and a parametrized state. Endo et al. \cite{Endo2019}, extended by McArdle et al. \cite{McArdle2018}, proposed calculating excited states of molecules using a variational method based on imaginary time evolution. More recently, Ollitrault et al. \cite{Ollitrault2019} implemented an extension of the subspace expansion method relying on quantum equation of motion to compute molecular excited energies on IBMQ.  \\ 
\indent While these could likely prove effective as quantum computers develop, they remain of limited use on NISQ devices as they require deep quantum circuits and/or a large number of measurements. For instance, the Quantum-subspace expansion methods is very sensitive to noise. Using a restricted Hilbert space expansion and diagonalization of noisy matrices can lead to systematic biases (see for example \cite{Blunt2018}). There are also known cases in which the classical analogue of the method fails \cite{Watson2012}. The method proposed in \cite{Higgott2019} is also challenging to implemented on NISQ devices as it either requires accurately learning the inverse circuit of each new state discovered, or utilising SWAP tests. SWAP gates, and as a result SWAP tests, have no known native implementations on superconducting or ion-trap QPUs. Their equivalent sequences in native gates are known to be particularly burdensome for accuracy \cite{Leymann2020, Cincio2018}. \\

\section{Method description} \label{sec:method}

\indent Our method is designed to require as little as possible from NISQ devices in order to focus on the earliest practical application of the technology. It relies on combining an orthogonality objective with an energy minimization objective (also named VQE objective). At a high level, the Discriminative VQE (DVQE) aims at finding a state orthogonal to the ground state which at the same time is at a minimum of the Hamiltonian energy landscape. This will correspond to an approximation of the first excited state: the Hylleraas-Undheim and MacDonald \cite{Hylleraas1930,MacDonald1933} theorem implies that the energy of a state orthogonal to the ground state (or any number of lower excitation states) acts as an upper bound for the next eigenvalue. \\
\indent Rather than directly minimizing the overlap of the excited state of interest with the previous excited states and/or the ground state (as is done for instance in \cite{Higgott2019}), our method uses a combination of two quantum circuits working in tandem to learn parametrization angles and reproduce unknown excited states. Our technique takes inspiration from Quantum Generative Adversarial Networks (QGAN). In a classical Generative Adversarial Network, an initial Generator network (denoted by $G$) is trained to fake an unknown data structure by learning how to fool a Discriminator network (denoted by $D$). The Discriminator is trained to distinguish between the generated data structure and the unknown data structure. The QGAN is an adaptation of this algorithm where the data structure is replaced by a pure quantum state. The parametrized quantum circuit is trained to generate an approximation of an unknown pure state \cite{Lloyd2018, Benedetti2019}. \\
\indent In our case however, the logic of the QGAN is reversed. Instead of trying to fool the Discriminator, the Generator learns to create a state which makes it as easy as possible for the Discriminator to distinguish between a known quantum state (for instance, a simulated ground state) and the generated state. In effect, the Generator is identical to the ansatz circuit used for the VQE, although with different parameters. Borrowing from the QGAN logic, one can see that this change would result in producing a state which is as easily distinguishable from the known state as possible. In classical problems, this approach rarely makes sense. In quantum problems however, a state which has no overlap with a given reference state will be in the latter's orthogonality space.\\
\indent There are an infinite number of physically meaningful orthogonal states to a given quantum state. The VQE objective is used to guide the learning of the Generator towards a single orthogonal state. A state which is orthogonal to the ground state and at the same time minimizes the energy of the entire orthogonal subspace must be the first excited state. \\
\indent With this in mind, we believe the method we propose offers the following advantages:\\
\begin{enumerate}
    \item It is decisively NISQ friendly, requiring only rotation gates, entangling gates, and only one additional qubit compared to a VQE.\\
    \item Our orthogonality objectives rely on single qubit measurements (as we use ancilla qubits), reducing exposure to read-out errors, and does not require computation of overlap terms which have been recognised as challenging for NISQ devices (\cite{Higgott2019}).\\
    \item The method used to enforce orthogonality does not require perfect optimization and is therefore quite resilient to quantum noise\\
    \item The excited state is directly and variationally minimized, rather than being inferred through non-linear postprocessing (as it is the case for example in analytic continuation of imaginary time or in subspace diagonalization). This in turn reduces exposure to systematic bias in the estimation result.\\
    \item Unlike some of methods outline above (in particular methods based on subspace expansion and its extention), the classical overhead is minimal and scales identically to the classical overhead of the VQE.
\end{enumerate}
\indent First consider a series of pure states $\rho_{s_{i}} {=} \Ket{s_{i}}\Bra{s_{i}}$, with $i {\in} [0, n]$ representing adequate approximations of the first $n$ excited states of a Hamiltonian $H$. It is assumed that we have a pre-trained quantum circuit that can produce these states using indexed parameters $\theta_i$ (which can be obtained using the VQE and previous iterations of this algorithm). We are looking for a way to determine the $n+1$ state: $\rho_{s_{n+1}} {=} \Ket{s_{n+1}}\Bra{s_{n+1}}$. \\
\indent For this, consider a state  $\rho_{g}$ generated through a parametrized quantum circuit applied to an initial state $\Ket{0}\Bra{0}^{\otimes d}$, and which is initiated as state $\rho_{n}$. We denoted this Generator circuit as $G(\theta)$, with parameters $\theta$. We have $\rho_{g}{=}G(\theta)\Ket{0}\Bra{0}G^\dagger(\theta)$.

    \begin{figure}[ht]
    \centering
         \includegraphics[width=1.\columnwidth]{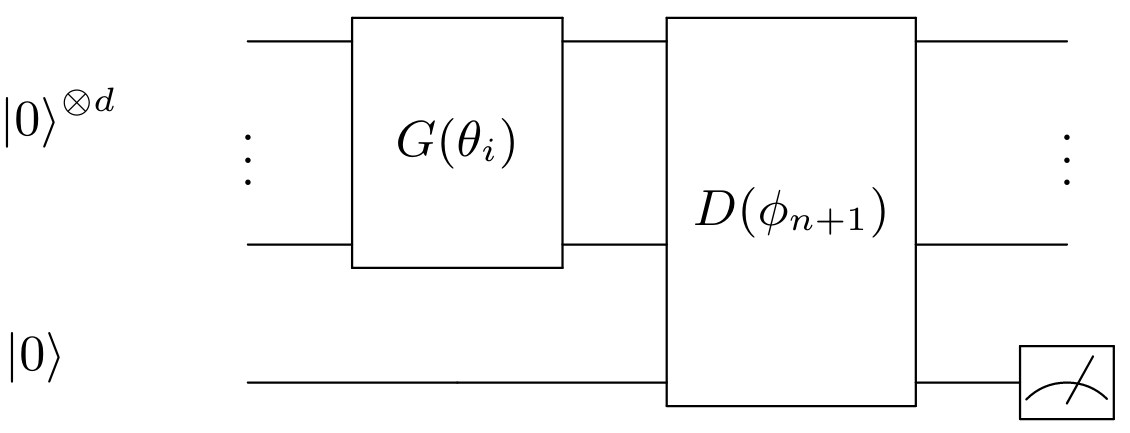}\par\medskip
    	 \caption{Discriminative VQE (DVQE) quantum circuit for computation of state $n + 1$. The hardware efficient ansatz is built using repeated layers composed of two rotation gates followed by entangling gates between nearest neighbour qubits, as described in \cite{Mitarai2019}}\par\medskip \label{fig:circuitnes}
    \end{figure}

\indent Consider a Discriminator quantum circuit labelled $D(\phi)$, which is tasked with distinguishing between any of the known states and the output of the Generator. In order to accomplish this task, it takes as input either any of the known states, or $\rho_{g}$, randomly but with equal probability. It is followed by a Positive Operator Value Measurement (POVM). Because of the discriminative objective of the circuit, we can limit the required POVM outcome to only two elements: 0 if the Discriminator identifies one of the known states, and 1 if it identifies the generated state. We can therefore map the POVM to a single ancilla qubit, also input to the Discriminator (see quantum circuit: FIG. \ref{fig:circuitnes}). We define $P_0$ as the projector of the circuit output state onto the zero state of the ancilla qubit.\\
\indent Based on this, and considering that the Generator cost function must also take into consideration the energy minimization objective we can define two subsequent cost functions that need to be minimized iteratively, for the Generator first and the Discriminator second. At optimum, the Generator cost function converges to the energy of the $n^{th}$ excited state: \\
	\begin{align}
		C^{(n)}_{gen}(\theta)  &= \Bra{0}G^\dagger(\theta)  H  G(\theta)\Ket{0} \nonumber \\
                         &+  \gamma Tr[P_{0}D(\phi)(\rho_{g}\otimes\Ket{0}\Bra{0})D^\dagger(\phi)],
	\end{align}
	\begin{align}
		C^{(n)}_{disc}(\phi)  &= Tr[P_{0}D(\phi)(\rho_{g}\otimes\Ket{0}\Bra{0})D^\dagger(\phi)]   \nonumber \\
			&-\sum_{n}Tr[P_{0}D(\phi)(\rho_{s_{i}}\otimes\Ket{0}\Bra{0})D^\dagger(\phi)].
	\end{align}
	
\indent We added a weighting factor $\gamma$ to the Generator cost function. This is to guarantee that the minimum of the optimization problem is indeed the state of index $n + 1$. For this, we must have $\gamma > (n + 1)(E_{n+1} - E_0)$. The derivation for the cost functions, convergence demonstration and explanation for the $\gamma$ factor can be found in the following sections (Sections \ref{sec:deriv} and \ref{sec:demons}, respectively). We can find a suitable $\gamma$ for all states by computing the maximum energy, running a VQE on the inverse Hamiltonian and taking the difference between the lowest energy state and the highest energy state. \\
\indent It is worth noting that the Generator cost function is identical for any excitation level, while a term is added to the Discriminator at each new level of excitation calculated (one for each level of excitation). Therefore, there is a linear increase in the number of terms to be calculated with the number of excited states. \\

\section{Derivation of the value functions} \label{sec:deriv}

\indent Consider an application of the DVQE circuit in which only the ancilla qubit is measured, The methodology to derive the DVQE value function is analogous to that developed for the QGAN in \cite{Benedetti2019}. We note $\rho_g$ the output state of the Generator and $\rho_{s_i}$ the excited state of index $i$. Similarly, we note the state of all qubits after the DVQE circuit $\rho_{Dg}$ for the generated state, and $\rho_{Ds_{i}}$ for any state $i \in [0, n]$. Recalling that we note $D(\phi)$ the operator resulting from the Discriminator circuit, $G(\theta)$ the operator resulting from the Generator circuit (we omit the $\phi$ and $\theta$ in our notations), that $\rho_g = G\Ket{0}\Bra{0}^{\otimes d}G^{\dagger}$, and that $\rho_{s_{i}}$ represent any known energy state of the molecule, we have:
	\begin{align}
		\rho_{Dg} = D(\rho_{g}\otimes\Ket{0}\Bra{0})D^\dagger,\\
		\rho_{Ds_{i}} = D(\rho_{s_{i}}\otimes\Ket{0}\Bra{0})D^\dagger,
	\end{align}  
where one can observe that we have now added an ancilla qubit, the necessity of which is explained later on.

\indent The Discriminator therefore outputs a mixture $\rho_{D_{mix}} = p(g)\rho_{D_g} + \sum_i p(s_i)\rho_{D_{s_i}}$, with $p(g)$ and $p(s_i)$ the probabilities of presenting the generated or any state $s_i$ to the Discriminator. We conduct a POVM on the output state, with projectors $P_b$, with $b$ indexing the possible measurement outcomes such that $\sum_b P_b = \mathbb{1}$. Each possible measurement outcome $P_{b}$, can occur with a probability $p(b) = Tr[P_{b} \rho_{mix}]$, following Born\textquotesingle s rule. The Discriminator can either be right and the POVM identifies correctly the input state, or the Discriminator can be wrong and the POVM identifies the incorrect input state. The process through which the POVM identifies the input state is refered to as the decision rules. \\
\indent Following Bayes' theorem, this decision rule should select the index $b$ which maximizes the posterior probability, or argmax$_{x\in \{g,s_i\}} p(x|b)$. It has been shown that this decision function (Bayes' decision function) has the lowest probability of error of any possible decision function \cite{Fuchs1996}.

\indent Our value function is built in order for the Discriminator to minimize the probability of error on a given measurement outcome. The probability of the measurement resulting in a correct decision is max$_{x \in \{g,s_i\}} p(x|b)$. Therefore, using Bayes' decision function, the probability of error when observing any element of the set $\{P_b\}$, can be written as:
	\begin{align}
		p_{err}(\{P_b\}) &= \sum_{b}(1 - \max_{x}p(x|b))p(b) \nonumber \\
			&= \sum_{b}\min_{x}p(x|b)p(b). \nonumber 
	\end{align}
\indent This equality is verified as the classification decision is done only over two possible categories: the Discriminator identifies a generated state $g$ or the Discriminator identifies any of the known states $s_i$. We therefore have $1 - \max_x p(x|b) = \min_x p(x|b)$. Given that by Bayes' formula $ p(x|b)p(b) = p(b|x)p(x)$:
	\begin{align}
 		p_{err}(\{P_b\}) &= \sum_{b}\min_{x}p(b|x)p(x) \nonumber \\
 			&= \sum_{b}\min_{x}Tr[P_{b}\rho_{x}]p(x).
	\end{align}
\indent The objective function for the Discriminator being to minimize the probability of error for any given outcome obtained, it can be described by
	\begin{equation}
		p_{err}^* = \min_{\{b\}}p_{err}(\{P_b\}),
	\end{equation}
where $\{P_b\}$ represents the set of projectors corresponding to all possible measurement outcomes.

\indent In our algorithm, we want the Discriminator to distinguish a generated state from any known state $\rho_{s_i}$. Therefore, the outcome of the POVM corresponds to the following: 0 is mapped to all the known states ($\rho_{s_{i}}$); 1 is mapped to the generated state ($\rho_{g}$).\\
\indent Noting $p(g)$ and $p(s_{i})$ the probabilities of the generated state and of any known state being presented to the Discriminator, the objective function is given by:
	\begin{align}
		&p_{err}^*  = \min_{\{P_{0}, P_{1}\}}(p(0|g)p(g) + \sum_{i}p(1|s_{i})p(s_{i}))           \nonumber \\
			&= \min_{\{P_{0}, P_{1}\}} (Tr[P_{0}\rho_{Dg}]p(g) + \sum_{i}Tr[P_{1}\rho_{Ds_{i}}]p(s_{i}))   \nonumber \\
 			&= \min_{\{P_{0}\}}(Tr[P_{0}\rho_{Dg}]p(g) + \sum_{i}Tr[(\mathbb{1} - P_{0})\rho_{Ds_{i}}]p(s_{i}))   \nonumber \\
			&= \min_{\{P_{0}\}}( Tr[P_{0}\rho_{Dg}]p(g) - \sum_{i}Tr[P_{0}\rho_{Ds_{i}}]p(s_{i})) + \sum_{i} p(s_{i}).
	\end{align}
\indent However this is also dependent on the action of the Generator. The objective of the Generator is $\min P_{err}^*$ w.r.t. $\rho_{g}$. Incorporating this objective in the equation above we get the following shared objective function:
 	\begin{align}
		& \min_{\{\rho_{g}\}}\min_{\{P_{0}\}}(Tr[P_{0}\rho_{Dg}]p(g) - \sum_{i}Tr[ P_{0}\rho_{Ds_{i}}]p(s_{i})) + \sum_{i} p(s_{i}).
	\end{align}
\indent Due to the discriminative objective of the circuit, we can limit the required POVM outcome to only two elements: 0 if the Discriminator identifies the original state, and 1 if it identifies the generated state. We can map the POVM to a single ancilla qubit also input to the Discriminator. In the case we have $P_b=\mathbb{1}^{\otimes d}\otimes \Ket{b}\Bra{b}, b \in [0, 1]$. Re-writing the state as the output of the quantum circuit we obtain the value function $\min_{\{\theta\}}\min_{\{\phi\}}V(\theta, \phi)$. Discarding the parametrization indices $\theta$ and $\phi$ we therefore aim to minimize
	\begin{align}
		V(\theta, \phi) &= Tr\left[P_{0}D(\rho_{g}\otimes\Ket{0}\Bra{0})D^\dagger\right] p(g) \nonumber \\
			&- \sum_{i=0}^n Tr\left[P_{0}D(\rho_{s_{i}}\otimes\Ket{0}\Bra{0})D^\dagger\right] p(s_{i})  \nonumber \\
			&+ \sum_{i=0}^n p(s_{i}).
	\end{align}
\indent The above value function is sufficient for the Generator to find at least one state belonging to the space orthogonal to all known states. However it does not guarantee that the state generated is $\rho_{s_{n + 1}}$. In order to do so, we can add a VQE objective to the value function, whereby the Generator will also aim at finding a state which then minimizes the expectation value of the Hamiltonian. Preemptively, we note that the weighting between both objectives is important in making sure the value function does converge to the desired excited state. In order to parametrize this weighting, we introduce a factor $\gamma$ the value of which is discussed in the following section. Re-writing the value function accordingly, we get
	\begin{align} \label{value_func}
		V(\theta, \phi) &= \Bra{0}G^\dagger  H  G\Ket{0} \nonumber \\
                                &+ \gamma\left[Tr[P_{0}D(\rho_{g}\otimes\Ket{0}\Bra{0})D^\dagger\right] p(g) \nonumber \\
			&-\gamma\sum_{i=0}^n Tr\left[P_{0}D(\rho_{s_{i}}\otimes\Ket{0}\Bra{0})D^\dagger\right] p(s_{i}) \nonumber \\
						&+ \gamma\sum_{i=0}^n p(s_{i}).
	\end{align}

\section{Convergence demonstration} \label{sec:demons}

Consider a generic state $\Ket{\psi} = \sum_{i=0}^{d-1}\alpha_i\Ket{s_i}$ such that $\Ket{\psi}= G(\theta)\Ket{0}$ (recalling that $d$ refers to the dimension of the system, and $n$ refers to the last excited state calculated). We use this state in the value function derived in equation \ref{value_func} (discarding $\theta$ and $\phi$ for readability):

    \begin{align}
        V &= \Bra{\psi}H\Ket{\psi}  \\ \nonumber
        & + \gamma Tr\left[P_0D(\Ket{\psi}\Bra{\psi}\otimes\Ket{0}\Bra{0})D^{\dagger}\right]p(g) \\ \nonumber
        & -  \gamma \sum_{i=0}^{n}Tr\left[P_0[D(\Ket{s_i}\Bra{s_i}\otimes\Ket{0}\Bra{0})D^{\dagger}\right]p(s_i) \\ \nonumber
        & + \gamma \sum_{i=0}^n p(s_i).
    \end{align}

The energy states $\Ket{s_i}$ form an eigenbasis for the molecular Hamiltonian which can be written in the form $H = \sum_i E_i\Ket{s_i}\Bra{s_i}$. We have $\Bra{s_i}H\Ket{s_i} = E_i$, and we can re-write the above equation as
    \begin{align}
        V &= \sum_{i=0}^{d-1}|\alpha_i|^2 E_i  \\ \nonumber
        & + \gamma Tr\left[P_0D(\sum_{i=0}^{d-1}\sum_{j=0}^{d-1}\alpha_i\alpha_j^{*}\Ket{s_i}\Bra{s_j}\otimes\Ket{0}\Bra{0})D^{\dagger}\right]p(g) \\ \nonumber
        & - \gamma \sum_{i=0}^{n}Tr\left[P_0D(\Ket{s_i}\Bra{s_i}\otimes\Ket{0}\Bra{0})D^{\dagger}\right]p(s_i) \\ \nonumber
        & + \gamma \sum_{i=0}^n p(s_i).
    \end{align}

\indent To simplify the writing, we set $p(g)$ and all $p(s_i)$ to be equiprobable, such that $p(g)=p(s_i)=\frac{1}{n + 1}$ (as we use $n$ known states $\Ket{s_i}$ plus the generated state) and $K_i=\frac{1}{n+1} Tr\left[P_0D(\Ket{s_i}\Bra{s_i}\otimes\Ket{0}\Bra{0})D^{\dagger}\right]$ and $k_i = \frac{1}{n+1} Tr\left[P_0D(\sum_{j\neq i}^{d-1}\alpha_i\alpha_j^{*}\Ket{s_i}\Bra{s_j}\otimes\Ket{0}\Bra{0})D^{\dagger}\right]$:

    \begin{align}
        V &= \sum_{i=0}^{d-1}|\alpha_i|^2 (E_i + \gamma K_i) + \sum_{i=0}^{d-1} \gamma k_i \\ \nonumber
        & - \gamma \sum_{i=0}^{n} K_i + \frac{n \gamma}{n+1}.
    \end{align}

\indent From here, we can see that the choice of set of parameters $\theta$, for the Generator, affect the values of the terms $\alpha_i$ (and therefore, also the terms $k_i$) while the choice of set of parameters $\phi$, for the Discriminator, affect the values of the terms $K_i$ and $k_i$. Both Generator and Discriminator are trained to minimize this value function, and it is clear that, as a result of the terms $k_i$, both need to be trained for a meaningful minimum to be found.\\
\indent It is important that the Discriminator is deep enough to be able to perform the classification between generated state and known states, we assume thereafter that it is the case. One can note that while some of the $K_i$ have both positive and negative factors in the value function (namely for $i \in [0, n]$), the $k_i$ all have positive factors. The terms $k_i$ should go to $0$ when the Discriminator is optimized. A similar argument can be made for the terms $K_i$ such that $i \in [n+1, d-1]$.  \\
\indent Here it is worth noting that these terms are in general not accessible to the user given the states $\Ket{s_i}$ for $i \in [n+1, d-1]$ are not known. However this does not prevent the convergence described above to occur during optimization.\\
\indent When the Generator is subsequently optimized, the value of the terms $k_i$ may increase as the $\alpha_i$ terms are updated. Subsequent updates of the Discriminator will bring these values back to $0$. This implies that Generator and Discriminator will need to be updated iteratively for the DVQE to work. To simplify the demonstration, we assume that the terms $k_i$ are sufficiently close to $0$ so that we can ignore them in the following. We have
    \begin{align} \label{simple_eq}
        V &= \sum_{i=0}^{d-1}|\alpha_i|^2 (E_i + \gamma K_i)
         - \gamma \sum_{i=0}^{n} K_i + \frac{n \gamma}{n+1}.
    \end{align}
\indent We now consider the case of optimizing the Generator in the context of Eq. \ref{simple_eq}, that is finding a minimum for this equation by only modifying the $\alpha_i$ terms and recalling that $\sum_i |\alpha_i|^2 = 1$. Because the terms $E_i + \gamma K_i$ can be ordered from smallest to largest, optimizing the Generator is equivalent to finding an index $p \in [0, d-1]$ such that $E_p + \gamma K_p < E_i + \gamma K_i $ for all $i \in [0, d-1] \setminus p$. In this case, $\alpha_p$ converges to $1$. \\
\indent In order to see that this index $p$ should equate to $n+1$ consider the ideal case in which the Discriminator is fully optimized and in which all $K_i$ with $i \in [0, n]$ are equal to $\frac{1}{n + 1}$. The last two terms in the Eq. \ref{simple_eq} cancel each other and we obtain a simplified value function
    \begin{align} \label{simple_eq2}
        V &= \sum_{i=0}^{d-1}|\alpha_i|^2 (E_i + \gamma K_i),
    \end{align}
which can be re-written as
    \begin{align} \label{simple_eq3}
        V &= \sum_{i=n+1}^{d-1}|\alpha_i|^2 (E_i + \gamma K_i) + \sum_{i=0}^{n}|\alpha_i|^2 (E_i + \gamma K_i).
    \end{align}
\indent Eq. (\ref{simple_eq3}) is important to understand how the algorithm behaves in a noisy environment, where the Discriminator cannot be fully optimized. However before discussing this, let us consider the case where the Discriminator perfectly succeeds at its task rendering $K_i = \frac{1}{n+1}$ for $i \in [0, n]$ and $K_i = 0$ for $i \in [n + 1, d - 1]$. We now have
    \begin{align} \label{simple_eq4}
        V &= \sum_{i=n+1}^{d-1}|\alpha_i|^2 (E_i) + \sum_{i=0}^{n}|\alpha_i|^2 (E_i + \frac{\gamma}{n+1}).
    \end{align}
\indent Once again, the action of optimizing the Generator will result in one of the $\alpha_i$ being equal to $1$, and the others to $0$. To make sure that it is $\alpha_{n+1}$ we must have $E_{n+1} < E_0 + \frac{\gamma}{n+1}$ or, the $\gamma$ factor, weighting the VQE and orthogonality objectives in the value function must obey 
    \begin{align} \label{gamma_eq}
        \gamma > (n + 1)(E_{n+1} - E_0).
    \end{align}

\indent In a more general case, considering equation \ref{simple_eq4}, for the state $n + 1$ to be the lowest energy of the value function, it must be that $(E_{n+1} + \gamma K_{n+1})$ is lower than $(E_{i} + \gamma K_{i})$ for any $i$ between $0$ and $d-1$ except $n + 1$. Therefore, given that together the Discriminator and the Generator push $K_i$ towards $0$ for $i$ greater than $n$ and towards $1$ for $i$ lower or equal to $n$ then it is possible for the algorithm to converge to the right state given a large enough $\gamma$ factor even if the Discriminator is not fully optimized. This is a particular advantage for NISQ computers where full optimization of the Discriminator and Generator may be impossible due to circuit and read-out errors creating an optimization barrier. \\ 
\indent We noticed however that in the case of a noisy QPU, using a $\gamma$ factor that is too high may result in the algorithm converging to the wrong value. That is because noise can prevent convergence to $0$ of the $k_i$ terms. If the Discriminator fails to bring close to $0$ the term $k_{n+1}$, it may be that the minimum of the value function is reached when more than one $\alpha$ term is non-zero. \\
\indent It is worth noting that the term $\frac{n \gamma}{n + 1}$ at the end of the value function has no impact on the optimization (as it has a null gradient in all parameters of the function). We could discard it and find the same optimal point. The value function at optimal point would be different but we would still find the eigenstate and eigenenergy. \\ 
\indent All together, by grouping the terms of the value function dependent on $\theta$ and the terms of the value function dependent on $\phi$, we find the cost functions of the Generator and of the Discriminator which have already been outlined in Section \ref{sec:method}: 

	\begin{align}
		C^{(n)}_{gen}(\theta)  &= \Bra{0}G^\dagger(\theta)  H  G(\theta)\Ket{0} \nonumber \\
                         &+  \gamma Tr[P_{0}D(\phi)(\rho_g\otimes\Ket{0}\Bra{0})D^\dagger(\phi)],
	\end{align}
	\begin{align}
		C^{(n)}_{disc}(\phi)  &= Tr[P_{0}D(\phi)(\rho_{g}\otimes\Ket{0}\Bra{0})D^\dagger(\phi)]   \nonumber \\
			&-\sum_{n}Tr[P_{0}D(\phi)(\rho_{s_{i}}\otimes\Ket{0}\Bra{0})D^\dagger(\phi))].
	\end{align}
	
\section{Experiments and results} \label{sec:results}

\subsection{Simulations}

\begin{figure}[ht]
	\centering
		\includegraphics[width=1.\columnwidth]{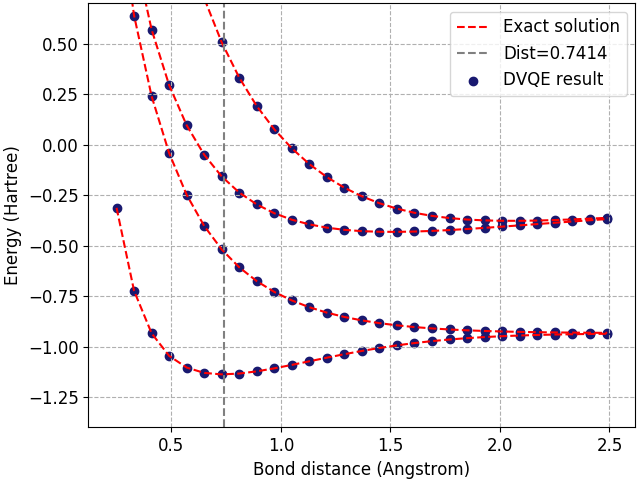}\par\medskip
	    \caption{Dissociation curves for $\mathrm{H_2}$  Hamiltonian using DVQE simulation and exact solver. Dotted lines represent ground and three excited states - all error to targets for this test are under 1 milliHartree}\par\medskip \label{fig:H2dissim}
	\textit{}
	\end{figure}

\indent In order to test our algorithm, we first simulated the excitation levels of the 2-qubit $\mathrm{H_2}$  Hamiltonian obtained using the Bravyi-Kitaev transformation in the STO-3G basis (results presented in FIG. \ref{fig:H2dissim}) We have used an optimization cycle of three iterations for the Discriminator followed by three iterations of the Generator, repeated iteratively until convergence. For this test, we use successive layers of the hardware efficient ansatz, as presented in \cite{Kandala2017}, each layer being composed of two rotations (one on the Y axis and one on the X axis) on each qubit, followed by a ladder of entangling gates. This results in a total of 8 parameters. The Discriminator is composed of three such layers (applied on 3 qubits and hence 18 parameters) for the first excited state and four such layers for the second and third excited state (hence 24 parameters). The algorithm first computes the ground state using the VQE and continues to determine the first excited state. Each subsequent excited state is computed iteratively once convergence has been reached on the previous one. Typically, a precision of $10^{-3}$ Hartree is achieved within $20$ iterations of the model using the Rprop optimizer \cite{Riedmiller1993}. \\
\indent We tested the algorithm on a 4-qubit version of the $\mathrm{LiH}$ Hamiltoninan, using the process detailed in \cite{Kandala2017} to build the Hamiltonians, and computing excitations until the $6^{th}$ excited state. We initially used a depth of four for the generator and of six to eight for the discriminator to model the ground state and the first three excited states of $\mathrm{LiH}$. Unlike the ansatz we used for $H_2$, we added rotations on the Z axis for each layers of the Hardware Efficient ansatz as it resulted in overall significantly better accuracy. We achieved a precision of at least $10^{-3}$ Hartree on average across bond distances for all excited states with maximum single error of $2.5$ milliHartree. This is offering an initial example to of the scalability of the method, showing precision is maintained on a larger system. To increase the expressiveness of the ansatz we added two layers to the generator to each subsequent energy state following the third state. Similarly, we increased depth by two layers for each subsequent energy state. While we know that the initial depth is not sufficient for computation of higher excited states, further research will be necessary to determine the optimal ansatz both for the generator and the discriminator.\\ 

	\begin{figure}[ht]
	\centering
		\includegraphics[width=1.\columnwidth]{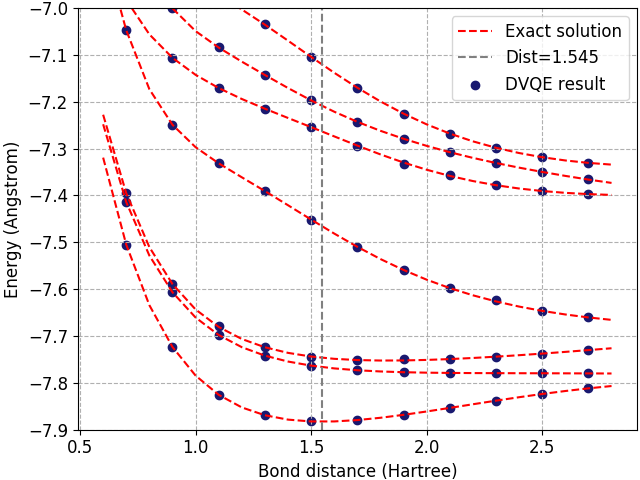}\par\medskip
	    \caption{Dissociation curves for $\mathrm{LiH}$ Hamiltonian using DVQE simulation and exact solver. Dotted lines represent ground and six excited states. Errors are on average below 1 milliHartree, with a few exceptions up to 2.5 milliHartree}\par\medskip \label{fig:LiHdissim}
	\textit{}
	\end{figure}
	
\subsection{QPU results}
\begin{figure}[ht]
	\centering
		\includegraphics[width=1.\columnwidth]{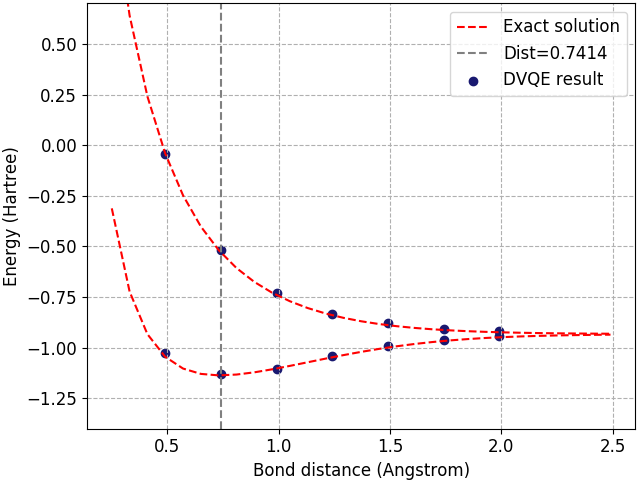}\par\medskip
	    \caption{Dissociation curves for $\mathrm{H_2}$  Hamiltonian using DVQE on IBMQ London, Ourense and Vigo and exact solver. Dotted lines represent ground and first excited state. Errors are within an average of 6 milliHartree for the ground state, and 8 milliHartree for the first excited state}\par\medskip \label{fig:H2dis_qpu}
	\textit{}
	\end{figure}

\indent In order to test the algorithm's resilience to errors, we implemented our algorithm on IBMQ London and Vigo Quantum Processing Units (QPUs) for the $\mathrm{H_2}$, two-qubit Hamiltonian (results presented in FIG. \ref{fig:H2dis_qpu}). Instead of using Rprop, we used the Rotosolve algorithm for which convergence is reached significantly faster \cite{Ostaszewski2019} at the expense of not being parallelizable. Read-out errors are mitigated using the IBMQ Qiskit Ignis tool (see Section \ref{sec:Implementation}). We computed both ground state through VQE and first state using DVQE. We found that both achieved about $10^{-2}$ Hartree accuracy. \\

\indent Computing second excited state would have required an additional layer for the Discriminator and as a result more involved error mitigation to obtain an accurate result. Similarly, higher accuracy would require stronger error mitigation methods or lower circuit error rates. In particular we estimated that, given the depth of circuits used and based on the data provided by IBMQ, our circuit error on runs of the Generator was about $2\%$ on all QPUs and of roughly $8\%$ on runs of the full DVQE (Generator plus Discriminator).\\

\section{Implementation details} \label{sec:Implementation}

\indent Running an algorithm on a QPU remains computationally costly. We focused on minimizing the number of single instruction requests to the QPU required to run the algorithm to an appropriate level of convergence. Each of our instruction requests covers the Rotosolve optimization of one angle for either the Generator or the Discriminator. It includes requests to conduct estimation (through a given number of measurements, or shots) of the three expectation value terms required to complete a Rotosolve iteration. \\
\indent Given the $\mathrm{H_2}$ Hamiltonian on two qubits, we used a circuit of depth 2 for the Generator and of depth 3 for the Discriminator, with each layer composed of two rotation gates ($R_X$ and $R_Y$) and an entangling gate. Hence we had to optimize 8 parameters for the Generator, and 18 for the Discriminator. The benefits of further depth could be studied but given our objective of minimizing the number of calls to the QPUs we have not attempted anything further outside of simulation. For each bond distance, we use 2 iterations of the Generator and 2 iterations of the Discriminator for each iteration of the DVQE, and a total of 4 iterations of the DVQE resulting in a total of 208 separate calls to the QPU for each point (in addition to what was required to compute the ground state, usually 2 iterations of the VQE, which has the same depth as the Generator, hence 18 calls). \\
\indent This optimization schedule was used only for calculating the energy values at a bond distance of $0.741$. For other bond distances, we performed a warm start by using the $\theta$ and $\phi$ parameters learnt at distance $0.741$ as a starting point for our optimization process. In all cases, one iteration of the VQE and one iteration of the DVQE was sufficient to reach convergence (although more were required to show convergence). In addition, it is worth noting that as the efficacy of the Discriminator is resilient to noise, it is also resilient to small changes in the bond distance. In particular, we noticed that we did not need to re-train the Discriminator in most cases in order to reach convergence. This however may not be true when studying more complex systems and when attempting to achieve higher accuracy (for instance by increasing the number of measurements beyond $8,000$). 
\indent In order to reduce the number of shots conducted, we used a ramping-up schedule for the circuit estimate. The first few iterations of the circuit are done with a low numbers of shots, and the final iteration of the DVQE was done using $8,000$ measurements. Energies are then calculated using the final $\theta$ obtained and using repeated $8,000$ shots run to obtain an average. \\
\indent It is worth noting that while we used Rotosolve for the implementation on a QPU, we used Rprop for the simulation. There are good reasons to think that this algorithm will be more relevant on a multi-core QPU than on a single QPU with a large number of qubits. Multi-core QPUs could offer tremendous opportunities for parallelization. Because calculations of angles under Rotosolve are co-dependent on each other, it offers less parallelization than gradient based methods such as Rprop where all angle gradients can be calculated in parallel. Whether Rotosolve or gradient based methods will be more efficient remains to be seen, however as long as QPUs are single core, Rotosolve will likely perform better for actual QPU runs, while Rprop (and other efficient gradient based methods) will be significantly more efficient for simulations.\\

\begin{table}

\begin{ruledtabular}
\begin{tabular}{cccccccc}
Bond distance &QPU& Energy &~~Exact~~&~~DVQE~~&\\
\hline
0.491 &Ourense &Ground &-1.047 & -1.025 \\
      &        &First  &-0.046 & -0.045 \\
0.741 &London  &Ground &-1.137 & -1.129 \\
      &        &First  &-0.532 & -0.519 \\
0.991 &Vigo    &Ground &-1.103 & -1.108 \\
      &        &First &-0.741 & -0.728 \\
1.241 &Vigo    &Ground &-1.048 & -1.040 \\
      &        &First &-0.840 & -0.832 \\
1.491 &Vigo    &Ground &-0.999 &-0.991 \\
      &        &First &-0.889 & -0.880 \\
1.741 &Vigo    &Ground &-0.967 &-0.964 \\
      &        &First &-0.913 &-0.908 \\
1.991 &Ourense &Ground &-0.949 &-0.943 \\
      &        &First &-0.924 &-0.919 \\

\end{tabular}
\end{ruledtabular}
\caption{\label{tab:table1}Detailed results of DVQE runs on IBMQ - values given are average of the last round of Rotosolve iteration (all in Hartree)}
\end{table}

\indent Errors on the measurement results were mitigated using the IBM Qiskit Ignis error mitigation tool. The process is described here briefly. We first measure the quantum computer prepared in each of the $2^n$ computational basis, where $n$ is the number of qubits. This could be easily achieved with quantum circuits using Pauli X gates and measurements. Using the measurement outcomes of the $2^n$ circuits, we could construct an estimate of the matrix $M$ defined element-wise as: \\
    \begin{align} \nonumber
        M_{i,j} &= \text{Probability}\{\text{measured state }i|\text{prepared in state j}\} \\ \nonumber
        &i,j\in\{0,1,\cdots 2^n-1\} \nonumber
    \end{align}
\indent Then, we would like to apply the inverse of $M$ to the measurement outcomes in the experiments. This is achieved by solving the following optimization problem: \\
    \begin{align} \nonumber
        x = \mathrm{argmin}_X |Y - M X|,\, \text{subject to} \sum_i X_i = \sum_i Y_i \nonumber
    \end{align}
where $Y$ is the vector of raw measurement outcome and $x$ is the vector of error mitigated measurement outcome. In the $i$th position of each vector is the number of occurrence of the measurement outcome in state $i$. The vector norm is defined as $|v| = v\cdot v$.\\ 
\indent The detailed results obtained are presented in table \ref{tab:table1}. The model was developed and tested using Hyrax. The simulations were conducted using a TensorFlow backend simulator, while the actual tests on QPUs used a Qiskit backend linked to IBMQ. 

\section{Error propagation analysis - using $\mathrm{LiH}$} \label{sec:error_prop}

    \begin{figure}[ht]
	\centering
		\includegraphics[width=1.\columnwidth]{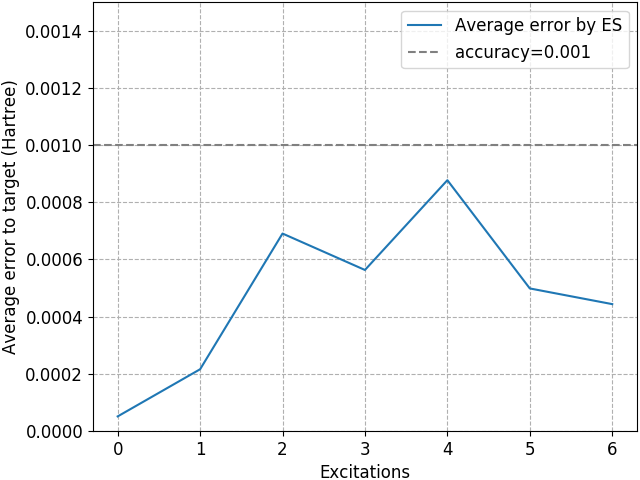}\par\medskip
	    \caption{Average absolute error by excitation level over bond distances 0.7 to 2.7 (step 0.2)}\par\medskip \label{fig:error_ave}
	\textit{}
	\end{figure}

\indent In this section, we study the accuracy of the DVQE simulation when applied to $\mathrm{LiH}$. On average, we find that convergence is reached within a $10^{-3}$ accuracy for all excitation levels with some outliers in higher excited states.  
Figure \ref{fig:error_ave} shows an increase in the magnitude of errors after ground and first excitation. It is worth nothing that we present average absolute error in this figure. In some cases, in particular for higher excited states, the DVQE converges slightly below the target value as a result of previous states not being perfectly orthogonal. These 'overshoots' errors however tend to be lower than 'undershoot' errors, resulting in higher excited states having better accuracy than some of the previous ones (e.g. second or fourth excited states).

	\begin{figure}[ht]
	\centering
		\includegraphics[width=1.0\columnwidth]{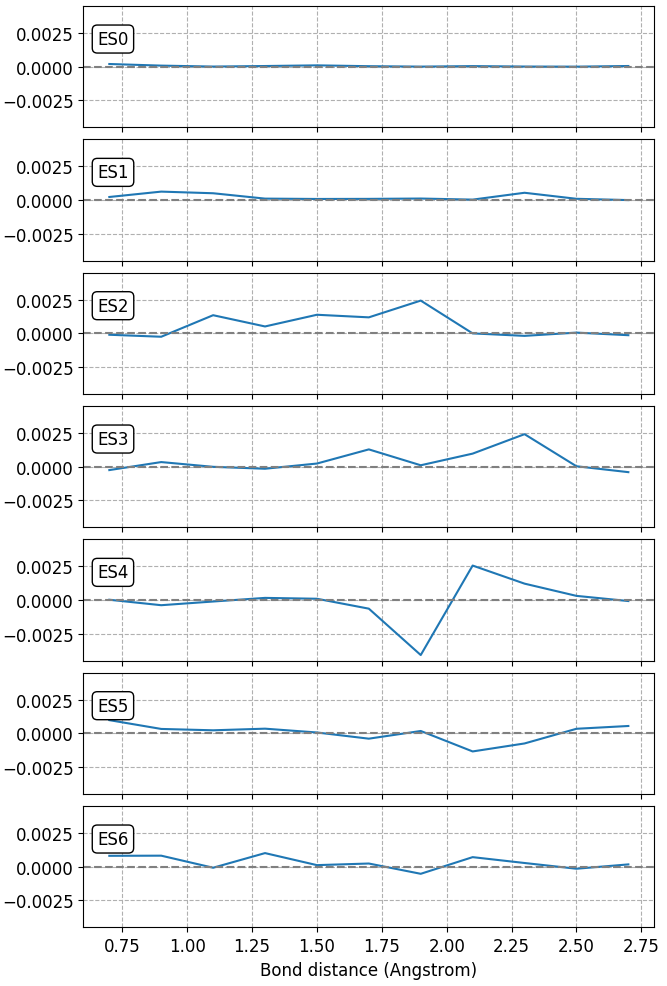}\par\medskip
        \caption{Complete set of errors by level of excitation and bond distance - Errors to target are in Hartree}\par\medskip \label{fig:errors_stack}
	\textit{}
	\end{figure}
	
Additionally, it appears that most of the reduction in accuracy in higher excited states is driven by a higher frequency of outliers (instances where accuracy is below $10^{-3}$). This is particularly visible when considering the third and fourth excited states in Figure \ref{fig:errors_stack}. Large outliers are systematically followed by low error overshoots in the following excited state. There are a number of reasons that could explains these outliers. First, this could be a result of our parameter initialisation strategy: we perform a warm start using the parameters of the nearest bond distance. While this reduces significantly the number of iterations to reach convergence, it could in some instances initialise the modelled wavefunction close to a local minimum, preventing convergence to the target value. Second, it could be that the ansatz is not expressive enough for certain bond distances (intuitively one can think that molecules with relatively higher bond distances have more entangled electrons). One factor that supports this second point is that we were able to increase average accuracy from the order of $10^{-2}$ to $10^{-3}$ Hartree at bond distance $2.3$ Angstrom by simply increasing the ansatz for the generator by one layer across the spectrum. \\ \\

\section{Conclusion}

\indent We have shown that one can find an accurate approximations of molecular energy spectra using the DVQE method, which can operate within the restrictions of NISQ devices. As for all the other excited states methods proposed for NISQ, scalability remains under question as gate errors remain too high to test much larger systems. Our results opens several avenues of research for modelling excited states on QPUs using fully variational methods. It highlights a number of research questions that remains to be addressed. 
\indent Further work will be required to determine an optimal ansatz structure both for the generator and the discriminator. In particular, the depth of the discriminator will likely be the bottleneck for any further implementation of this algorithm on QPUs. As we have seen, additional depth required for the second excited state of $H_2$ renders it too deep to be implemented reliably at this stage. This echoes to the need for further improvements in error mitigation techniques, especially with regards to extrapolation ~\cite{Li2017, Temme2017, Endo2018_mit} which will be critical for producing valuable computation on NISQ devices. 

\bigskip

\indent We would like to thank Pr. Jonathan Tennyson, Dr. George Booth and Dr. Thomas Rogers for their detailed feedback and advice, and Dr. Antonio Mezzacapo for providing Hamiltonian data computed in \cite{Kandala2017}. JT is supported by the UK EPSRC [EP/R513143/1]. HC acknowledges support through a Teaching Fellowship from UCL. LW acknowledges support through the Google PhD Fellowship in Quantum Computing. EG is supported by the UK EPSRC [EP/P510270/1].

\providecommand{\noopsort}[1]{}\providecommand{\singleletter}[1]{#1}%

\bigskip

\end{document}